\documentclass[sigconf]{acmart}
\usepackage{verbatim}
\usepackage{subcaption}
\usepackage{color}
\usepackage{xcolor}
\usepackage{multirow}
\AtBeginDocument{%
  \providecommand\BibTeX{{%
    \normalfont B\kern-0.5em{\scshape i\kern-0.25em b}\kern-0.8em\TeX}}}

\copyrightyear{2024}
\acmYear{2024}
\setcopyright{acmlicensed}\acmConference[CUI '24]{ACM Conversational User Interfaces 2024}{July 8--10, 2024}{Luxembourg, Luxembourg}
\acmBooktitle{ACM Conversational User Interfaces 2024 (CUI '24), July 8--10, 2024, Luxembourg, Luxembourg}
\acmDOI{10.1145/3640794.3665545}
\acmISBN{979-8-4007-0511-3/24/07}

\begin{document}

\title{The Impact of Perceived Tone, Age, and Gender on Voice Assistant Persuasiveness in the Context of Product Recommendations}

\author{Sabid Bin Habib Pias}
\affiliation{
  \institution{Indiana University}
  \city{Bloomington}
  \country{USA}
}
\author{Ran Huang}
\affiliation{
  \institution{Indiana University}
  \city{Bloomington}
  \country{USA}
}
\author{Donald Williamson}
\affiliation{
  \institution{Ohio State University}
  \city{Columbus}
  \country{USA}
}
\author{Minjeong Kim}
\affiliation{
  \institution{Indiana University}
  \city{Bloomington}
  \country{USA}
}
\author{Apu Kapadia}
\affiliation{
  \institution{Indiana University}
  \city{Bloomington}
  \country{USA}
}

\renewcommand{\shortauthors}{Sabid Bin Habib Pias et al.}
\renewcommand{\shorttitle}{Impact of Perceived Tone, Age, and Gender on Voice Assistant Persuasiveness}
\newcommand{\quickwordcount}[1]{%
  \immediate\write18{texcount -1 -sum -merge -q #1.tex output.bbl > #1-words.sum }%
  \input{#1-words.sum} words%
}
\newcommand{\detailtexcount}[1]{%
  \immediate\write18{texcount -merge -sum -q #1.tex output.bbl > #1.wcdetail }%
  \verbatiminput{#1.wcdetail}%
}


\begin{abstract}
  Voice Assistants (VAs) can assist users in various everyday tasks, but many users are reluctant to rely on VAs for intricate tasks like online shopping. This study aims to examine whether the vocal characteristics of VAs can serve as an effective tool to persuade users and increase user engagement with VAs in online shopping. Prior studies have demonstrated that the perceived tone, age, and gender of a voice influence the perceived persuasiveness of the speaker in interpersonal interactions. Furthermore, persuasion in product communication has been shown to affect purchase decisions in online shopping.  We investigate whether variations in a VA voice's perceived tone, age, and gender characteristics can persuade users and ultimately affect their purchase decisions. Our experimental study showed that participants were more persuaded to make purchase decisions by VA voices having positive or neutral tones as well as middle-aged male or younger female voices. Our results suggest that VA designers should offer users the ability to easily customize VA voices with a range of tones, ages, and genders. This customization can enhance user comfort and enjoyment, potentially leading to higher engagement with VAs. Additionally, we discuss the boundaries of ethical persuasion, emphasizing the importance of safeguarding users' interests against unwarranted manipulation.

\end{abstract}



\maketitle

\section{Introduction}
Voice assistants (VA) like Amazon's Alexa, Google Assistant, and Apple's Siri~\cite{AmazonAlexa, GoogleAssistant, AppleSiri} have become incredibly popular due to their hands-free convenience and wide range of applications~\cite{vausage, alexabuy}. Popular functionalities of voice assistants encompass information seeking, including weather checks, recent news, or cooking instruction~\cite{luger2016like, lopatovska2019talk, 10.1145/3613904.3642183}, entertainment, such as music or jokes, and controlling external devices~\cite{lopatovska2019talk}. Because of recent advancements in natural language processing (NLP)~\cite{pinsky2021loud} through the use of large language models (LLMs), VAs have vastly expanded their capabilities to perform intricate tasks, such as engaging in a conversation with users or offering online shopping recommendations~\cite{de2020reducing, alexashopping}. Furthermore, VA recommendations have demonstrated greater efficacy compared to text-based recommendations~\cite{flavian2023effects}, making them a suitable alternative for online shopping guidance. However, people are less comfortable with AI assistance in risky tasks that may have financial consequences compared to tasks with low risks~\cite{novozhilova_looking_2024, vaownernerve}. Evidence suggests that the perceived humanlike characteristics of a machine increase users' perceived comfort~\cite{kamide_perceived_2017}. Moreover, Li and Sung posited that anthropomorphism reduces the psychological distance in human-AI assistant interactions~\cite{ li2021anthropomorphism}. Therefore, improving the social appearance of a VA can be a stepping stone toward increasing the adoption of intricate tasks in VAs.

To increase anthropomorphism in VAs, researchers have explored the effect of human-like vocal characteristics in VAs~\cite{wagner2019alexa, doyle2019mapping, wagner2019human, phinnemore2023creepy}. People perceive news stories as more interesting if the synthesized voice expresses them in a happier voice, as opposed to a sad tone~\cite{nass2001effects}. Additionally, Belanche et al. demonstrated that integrating perceived warmth into a service robot's voice increased user expectations regarding the quality of the robot's service~\cite{belanche2021examining} and studies on interaction with synthesized voices found that users generally prefer female and extroverted VA voices~\cite{ernst2020impact}. Furthermore, users generally prefer VAs to have a friendly and engaging tone rather than sounding robotic or monotone~\cite{guha2023artificiality}. However, the implications of differing tones, ages, and genders of VA voices on the perceived persuasiveness of VAs and users' purchase decisions remain largely unexplored.

Current voice assistants let users personalize how they sound to some extent---Amazon's Alexa~\cite{AmazonAlexa} offers one male and one female voice each for a limited number of accents, such as US or UK accents. In addition, Alexa lets users adjust the speaking rate or switch to a whisper mode~\cite{alexawhisper}. Google Assistant provides 12 different voices, including male and female voices in varied accents~\cite{googlecolor}. Siri also offers a few different male or female voices in various accents~\cite{sirivoice}. However, these voice assistants offer limited voice options, with minimal diversity in vocal age and tone for user preference.

Besides linguistic variation, paralinguistic traits, such as the tone and the perceived age and gender of the voice, are effective in building trust in interpersonal relationships~\cite{glaser2016conversational}. Evidence suggests that pitch variance, speaker’s age, and fluency persuade listeners~\cite{belin2019sound, schirmer2020angry}. Furthermore, being perceived as entertaining and interesting positively influences the adoption and retention of voice assistants among users~\cite{zhong2022user}. Like interpersonal relationships, users likely have individual preferences in conversational tones, and thus, respond to various voices differently, whether consciously or unconsciously~\cite{krenn2017speak, dahlback2007similarity, dahlback2001spoken}. Therefore, variations in tone, age, or gender of the VA voice have the potential to influence users' perceptions of voice assistants in critical decision-making contexts. If making VAs sound different in tone, age, or gender helps users feel more at ease when they make critical decisions such as online shopping, then VA designers can utilize different voice attributes to broaden the adoption of VAs in critical contexts. Moreover, such flexibility will facilitate research exploring variations in VA tasks within analogous contexts such as banking and bill payments. Enhancing trust and comfort in using VAs for such intricate tasks can encourage more users to adopt VAs for intricate daily activities, paving the way to make individuals' daily lives easier through hands-free alternatives, particularly benefiting elderly individuals or  those with disabilities or visual impairments. However, ensuring healthy persuasion is vital to prevent unwanted manipulation and to protect the users' interests. VA designs should follow ethical guidelines that prioritize user welfare and consent. In addition, clear disclosure of persuasive intent and flexibility to control the vocal characteristics are also critically important in VA design.

The interplay between perceived age, tone, and gender of VAs in shaping persuasiveness and purchase decisions is yet to be explored. To address the gap in current literature, we are particularly interested in evaluating how the perceived tone, age, and gender of a VA's voice influence users' perceptions of persuasiveness and their purchase decisions.  In particular, we sought answers to the following questions: 

\textit{\textbf{RQ1:}} ``How does the perceived tone of a voice assistant's voice persuade participants and subsequently affect participants' purchase decisions?'' We aimed to explore how positive, negative, and neutral tones used by VAs persuade participants, and consequently, whether the participants follow VA suggestions for purchase decisions. Our objective was to examine the effect of changing tones regardless of the age or gender of the voice.

\textit{\textbf{RQ2:}} ``How do the perceived gender and age of a voice assistant's voice persuade participants and subsequently affect participants' purchase decisions?'' We investigated whether the persuasiveness of middle-aged and younger adult male/female voices in voice assistants are different and whether the persuasiveness affects participants' purchase decisions.

To investigate these questions, we developed an online experimental study with synthesized voices to examine how the perceived tone, age group, and gender of a VA's voice persuade participants and how the persuasiveness affects participants' purchase intentions. We also collected open-ended responses from the participants to further investigate their preferences toward particular voice types.

Our results showed that positive and neutral tones significantly persuaded the participants. As a result of the persuasion, the participants were more likely to follow VA recommendations in purchase decisions for positive and neutral tones. Interestingly, middle-aged male and younger female voices significantly persuaded the users and showed a similar influence on participants' purchase decisions.

With the advancement of voice assistants that require increased user engagement, the findings of this research can help subsequent research endeavors. Diversifying the tones, age, and gender of VA voices can effectively increase user involvement in intricate tasks with financial risks through healthy persuasion. However, an ethical framework to govern the use of persuasion and effective disclosure of potential persuasion is also important to safeguard users from unexpected manipulation.  Through responsible VA design, future VAs can enhance user involvement in intricate tasks, fostering a closer connection between users and VAs.

\section{Related Works}
In this section, we discuss the literature on a) user behavior in online shopping with voice assistants, and b) approaches to understanding the interplay between VA voice persuasiveness and the tone, age, and gender of the VA voice in interpersonal and human-computer interaction (HCI) contexts. 

\subsection{Decision Making for Online Shopping with Voice Assistants}
VA platforms offer decision-making assistance in online shopping through conversations~\cite{perez_voice_2018, tassiello_alexa_2021}. The conversations provide options including looking up suitable products and browsing customer reviews~\cite{mari_evolution_2020, ma_challenges_2020}. However, the perceived risk of negative consequences~\cite{campbell_moderating_2001, li_analysis_2009} and the lack of controllability of VAs challenge the wide adoption of online shopping with VAs. Rzepka et al. demonstrated that users find VAs enjoyable and convenient for general tasks, but are perceived as risky because of the perceived lack of reliability, opacity, and controllability of VAs in online shopping~\cite{rzepka_voice_2023}. Muthukumaran and Vani posited that consumers are hesitant to use VAs for shopping because of the possible risk of payment malfunction~\cite{muthukumaran_optimizing_2020}. Hong et al. showed that uncertainty with the technology causes user anxiety when online shopping with VAs~\cite{hong_what_2020}. Perceived risk in adoption of a new technology has been observed in internet banking~\cite{lee_factors_2009} and similar high involvement contexts~\cite{torki_biucky_effects_2017, zaharia_voice_2021}. 

However, increasingly human-like behavior by the VAs can ameliorate users' involvement in online shopping with VAs. Lee et al. showed that social presence of the VA helps increase the perceived usefulness of VAs in an online shopping context. As a result, users ultimately show more intention to use VAs for online shopping~\cite{lee_adopting_2023}. Dellaert suggested that human-like communication by VAs increases users' online shopping activity with VAs~\cite{dellaert_consumer_2020}. Clausen et al. found humorous responses by VAs effective in recovering mistakes~\cite{clausen_exploring_2023}. These findings support the implication of the `computers are social actors' (CASA) paradigm~\cite{nass1994computers} in an online shopping context with VAs.
The CASA paradigm posits that social cues in HCI can potentially trigger similar social behaviors involving real-life scenarios. In this context, users react to human-like activities of VAs as they would react to a human being. For instance, Rhee and Choi demonstrated a positive effect of a VA with friendly language in a voice shopping context~\cite{rhee2020effects}. Zhu et al. showed that users liked voice chatbots when the chatbots displayed emotional expressiveness~\cite{zhu_effects_2022}. Similarly, McLean et al. demonstrated the importance of social presence, perceived intelligence, and social attraction in building user engagement in online shopping~\cite{mclean_alexa_2021}. Lee et al. showed that perceived emotion toward VAs can strengthen users' mental involvement with the VA~\cite{lee_where_2022}. In contrast, there are opposing views on integrating human-like attributes into conversational agents in recommendation tasks. Starke and Lee argued that the quality of conversational aspects of a conversational recommender system might be slightly lowered if the recommender does an appropriate job in recommending a personalized item~\cite{starke_unifying_2022}. Moreover, Roesler et al. demonstrated that people prefer lower anthropomorphism in robots for specific contexts, such as industrial domains~\cite{roesler_why_2022}.

There are various scholarly perspectives on whether designing VAs with human-like characteristics may enhance user comfort and consequently increase the likelihood of users adhering to the VAs' recommendations. This diversity of viewpoints has sparked our interest in investigating the impact of vocal attributes on the persuasiveness of VAs, with a specific focus on determining whether such characteristics passively influence users' inclination to follow VA suggestions in the context of online shopping.

\subsection{Voice Characteristics and Persuasiveness}

Numerous studies have explored persuasiveness as a way to influence users' shopping outcomes in online shopping contexts ~\cite{adaji_e-commerce_2020, maslowska_effect_2017}. Persuasion is the process of influencing a person's attitude or behavior directly or indirectly by some form(s) of action~\cite{eagly1984cognitive, o2015persuasion}. Peripheral cues including trustworthiness, attractiveness, and tone~\cite{fogg2002persuasive, rocklage2018persuasion, li2021anthropomorphism, poyatos1993paralanguage, marrero2022evaluating} can influence the outcome of persuasion in interpersonal interaction. For instance, Wirz et al. found that perceived emotion in a political speech persuades the audience~\cite{wirz2018persuasion}. Moreover, confidence expressed by a human voice was shown to effectively enhance persuasion among audience members~\cite{van2020voice}. 

A similar effect of vocal characteristics on user perception has also been observed in the human-computer interaction domain. Kessens et al. designed a robot to educate and motivate children and showed that emotional vocal and visual cues effectively persuaded the children~\cite{kessens2009facial}. Similarly, people perceive news stories as more interesting if the synthesized voice expresses them in a happier voice, as opposed to a sad tone~\cite{nass2001effects}. Additionally, Belanche et al. demonstrated that integrating perceived warmth into a service robot's voice increased user expectations regarding the quality of the robot's service~\cite{belanche2021examining} and studies on interaction with synthesized voices found that users generally prefer female and extroverted VA voices~\cite{ernst2020impact}. Kim et al. found that urgency in a voice agent's tone increased user trust in emergency situations~\cite{kim_urgency_2023}. Dubiel et al. showed that people consider synthetic voices more truthful when they have a debating tone compared to a story-telling tone~\cite{dubiel2020persuasive}. Torre et al. observed that including the sound of a smile in the recommendation of a voice agent elicited higher trust in an investment gaming context~\cite{torre2020if}. 
Papenmeier and Topp demonstrated that meaningful acknowledgment from conversational agents increases the perceived competence of the agent~\cite{papenmeier_ah_2023}.

For VAs, studies have largely explored how the persuasiveness and impression of voice agents can be improved with linguistic or visual expressions~\cite{dubiel2022conversational, herder2023context, ischen2022voice, liebrecht2021linguistic, 10.1145/3613905.3650918}. Linguistic improvisations including shorter and longer responses~\cite{ gretry2017don, javornik2020don}, and informal and formal responses~\cite{johnen2019pushing, zhao2019appealing} have demonstrated persuasiveness in diverse conversational agent contexts. Similarly, there is an extensive body of research on user perceptions of synthesized voice attributes including tone, age, and gender~\cite{skoog2015can, sataloff2020effects, baird2018perception, baird2017perception}. However, there is a dearth of research specifically examining the interplay of tone, age, and gender in VAs' voices concerning their perceived persuasiveness. Our research addresses this gap, probing the associations between the persuasiveness of VAs and the tone, age, and gender of their voices, and subsequently how the persuasiveness leads users to consider VAs' suggestions. Previous studies have explored the impact of positive and negative words uttered by the user to increase the effectiveness of conversational interfaces' recommendations for online fashion choice~\cite{wu_multimodal_2022}. We have taken a similar approach, but for the tone of the VA voice, selecting a positive and negative tone of the VA voice.

Overall, the vocal attributes of tone, age, and gender collectively contribute to how a speaker is perceived~\cite{schuller1988emotion, gender2017paralinguistic}. Hence, these attributes can play a pivotal role in shaping users' perceptions of the persuasiveness of VAs. Therefore, our study emphasizes select vocal characteristics---tone, age, and gender---to examine the perceived persuasiveness of voice assistants.

\section{Method}
First, we generated voice stimuli with varying tones, age groups, and genders. Next, we validated the voice stimuli and measured the user behavior for varying voices of the VA. In the following subsections, we discuss the components of the study.

\subsection{Stimuli Generation}
We used Microsoft speech studio\footnote{\url{https://speech.microsoft.com/}} to generate voices that varied by tone, age group, and gender following the age group and gender criteria outlined by Waller et al.~\cite{skoog2015can}. We selected female and male voices and age groups including younger adults (20--30 years old), middle-aged adults (40--50 years old), and older adults (60--70 years old). We applied positive, neutral, and negative tones to each voice category. In Microsoft Speech Studio, we used a few template voices with emotional tones. To avoid complexity, we categorized these into three tones---positive tones including happy, excited, and cheerful; neutral, reflecting the default or normal and flat tone; and negative tones encompassing sad and frustrated tones. In addition, we varied the pitch and speech rate (Table~\ref{tab:pitch-sr}) of the voices to create different age group impressions, based on previous research showing that listeners perceive higher-pitched voices as younger and lower-pitched voices as older~\cite{winkler2007influences}. We edited the recordings in Audacity\footnote{\url{https://www.audacityteam.org/}} to normalize for intensity~\cite{skoog2015can}.

We used positive and negative product reviews from the popular online shopping platform Amazon\footnote{\url{https://www.amazon.com/}} and chose products from the most popular product categories purchased in 2023: Home \& Kitchen, Office Supplies, and Bags \& Luggage~\cite{statistamostpurchase, bluecartmostpurchase, junglescoutmostpurchase}. We selected gender-neutral products to avoid gender bias toward a product, including categories such as tumblers, desks, chairs, etc. Positive product reviews were generated with either positive or neutral tones and negative product reviews were generated with either negative or neutral tones. We excluded negative tone for positive reviews and positive tone for negative reviews, as the cross-valence effect was beyond the scope of our research. Overall, we generated 24 combinations of voices (two genders x three age groups x two review valences x two tone valences for each review valence). The reviews contained 26 to 33 words, aligned to the text length of the speech materials by Waller et al.~\cite{skoog2015can}. Thereby, the corresponding audio clips were between 12 and 16 seconds.

Following several iterations of voice synthesis, we preselected a group of voices that appeared to be appropriate for the corresponding target age groups and tones. Subsequently, we conducted a stimulus validation study to ensure that the voices were perceived as intended in terms of age group and tone.

\subsection{Stimuli Validation}
A preliminary study was conducted to identify the perceived tone and age group of the voices (N = 78). We adapted the voice perception test for assessing the perceived age and gender of synthesized voices developed by Baird et al.~\cite{baird2017perception, baird2018perception}.

We conducted several pilot studies and found that the participants faced difficulty in detecting the age of the older adult voices. Therefore, we removed the older adult voices from the stimulus set. In the final stimuli validation study, we used 48 audio clips containing 16 different combinations of the voices (two genders x two age groups x two review valences x two tones for each review valence) and each combination had three different voices. 

Each participant was presented with 14 audio clips including two attention check audio clips. We paid all the participants regardless of their attention check scores, but we used only the responses of the participants with zero attention check failure. We randomly presented only male or female voices to a participant so that the participants did not have a direct bias toward a particular gender, with 38 listening to male voices and 41 to female voices. For all audio clips, the participants responded to questions about their perception of the tone of the voice (seven-point Likert scale: strongly negative = -3 to strongly positive = 3)~\cite{amon2020influencing} and age-group of the voice (younger adult 20--30, middle-aged adult 40--50). The participants were able to play the audio multiple times and could stop, pause, and resume the audio. We included a test audio to let the participants adjust the volume of the audio on their devices. We calculated the average ratings across the participants for each voice. Each voice was then categorized into negative, neutral, or positive tones based on the tone ratings, and younger adults or middle-aged adults based on the frequency of age-group selection. We paid \$3.50 per participant according to the minimum wage guidance in the study location~\cite{silberman2018responsible}. Participant demographics are detailed in Appendix Table~\ref{tab:demography}.
 
\subsection{User Behavior Study}

\subsubsection{Stimulus and Procedure.} This experiment investigated user perceptions of varying vocal characteristics of a voice assistant in the context of online shopping and was conducted using the online survey platform Prolific\footnote{\url{https://app.prolific.co/}} in line with similar studies~\cite{tolmeijer2021female, huang2023should}. Participants (N = 335) were given a scenario to imagine themselves searching for a certain product and checking product reviews using a voice assistant. Each participant was presented with eight audio clips including two attention check clips. To avoid an idiosyncratic effect of a specific product and enhance the generalizability of our findings beyond a single product~\cite{fontenelle1985generalizing}, we used stimulus sampling of the reviews from six different household products such as dishwasher soap or hand soap, for example.

We conducted the study with a 2 (review valence: positive vs. negative) x 2 (voice assistant gender: female vs. male) x 2 (voice assistant age: younger adult vs. middle-aged adult) x 3 (voice assistant tone: positive/negative vs. neutral) between-subjects factorial design. Hence, one participant encountered only one voice type. The between-factor design for all variables was done to avoid any carry-over effect. In total, there were 96 audio clips for six products, with 16 clips each (eight positive and eight negative reviews for each product). After adjusting the device volume in an audio test, participants listened to the audio reviews one at a time. These reviews contained 26 to 35 words and the audio clips ranged from 12 seconds to 17 seconds. For each audio clip, we measured participants' purchase likelihood of the corresponding product (\textit{I am likely to purchase this product}~\cite{williams2022third}).

The following questions focused on participants' perceptions of the simulated VA and the VA voice. We measured participants' perceived persuasiveness of the voice (Table~\ref{tab:ques-voice}) alongside the perceived trustworthiness, usefulness, and enjoyment of the VA. The participants responded to these questions only once during the study. 

\begin{table}[ht]
  \begin{tabular}{ccc}
        \hline
      Factor & Item & Chron.\\
     \hline
      \multirow{2}{*}{Persuasiveness~\cite{pham2004ideals}} & The voice is compelling & \multirow{2}{*}{0.89} \\
      & The voice is convincing &   \\
       \hline
  \end{tabular}
  \caption{Question Items about Voice Persuasiveness. Chron. = Chronbach's alpha}
  \label{tab:ques-voice}
\end{table}

Next, we presented the participants with two qualitative open-ended questions:

\begin{itemize}
    \item What do you like/dislike about the voice you heard during the survey? Please elaborate.
    \item If you could create a voice for the voice assistant, what would that voice sound like? Please elaborate.
\end{itemize}

The remaining questions focused on VA usage, disposition of trust, and demographics (Appendix Tables~\ref{tab:ques-sva-usage-trust}). 

\subsubsection{Participants.} 
We adopted the a priori G*Power analysis~\cite{faul2007g} by Ischen et al.~\cite{ischen2022voice} for a between-subjects one-way analysis of variance (ANOVA) with four groups and found a required sample size of 320. We invited a total of 350 participants in Prolific and the responses of 335 without attention check failure were used for further analyses. More than 94\% participants had experience using a voice assistant (Appendix Table~\ref{tab:demography}). In this study, 75\% of the participants completed the survey in 17 minutes or less and were paid \$5 to conform with the minimum wage recommendation in the study location~\cite{silberman2018responsible}. Of the 335 eligible participants, 168 were presented with female voices and 167 listened to male voices. Furthermore, 159 participants listened to middle-aged adult voices and 176 heard younger adult voices. In terms of tones, voices considered positive, negative, and neutral voices were distributed to 80, 75, and 180 participants respectively.

\subsection{Quantitative Analysis}

We used the causal mediation analysis framework~\cite{baron_moderatormediator_1986, tingley_mediation_2014} in R using the package `lavaan'~\cite{rosseel_lavaan_2012} to estimate the direct effect of the perceived tone, age, and gender of the VA voice on the purchase decision of the participants and the indirect effect through the perceived persuasiveness of the VA voice. We chose the best mediation model based on the comparative fit index (CFI)~\cite{bentler_comparative_1990} and significant model test statistics. In addition, we conducted linear regression to evaluate the relationship between VA voice persuasiveness and VA vocal characteristics, i.e., tone, age, and gender of the VA voice. The data we collected met the assumptions of homoscedasticity and normality of the residuals. 

The purchase likelihood score was reversed into `purchase decision' for negative reviews to represent the influence of the VA in the purchase decision variable. We treated VA voice persuasiveness and purchase decision as dependent variables. We had four between-subject independent variables (gender, age group, tone of voice, and review valence); the independent variables were encoded as categorical variables. In all models, the participant ID was encoded as a random categorical variable. For the dependent variables with multiple items in the dataset, we used Chronbach's alpha~\cite{cronbach1951coefficient} to measure the internal consistency among the items in a group and we took the mean of the items to represent the corresponding dependent variable. In addition, we used Chronbach's alpha to measure the internal consistency among the responses for the six products (Cronbach's alpha~-- 0.96). We used estimated marginal means to compute the pairwise comparisons following the regression analysis to help make sense of the significant effects across the interaction effects. We adjusted the p-values using the Tukey method~\cite{williams1999controlling} as it considers multiple comparisons and adjusts the p-value to minimize the risk of Type I errors.

\subsection{Qualitative Analysis}

We used an inductive (thematic) analysis process for analyzing the open-ended questions about liking or disliking the voices and participants' preferences for the VA voice. We chose to perform thematic analysis because we did not derive the codes from any theory or prior research and our purpose was not to build new theories in the work. We used Delve\footnote{\url{https://delvetool.com/}} for coding the responses and identifying the themes. Initially, the researchers developed a code book for codes in an iterative bottom-up approach by weekly meetings. Then, based on the code book, we coded the rest of the responses. From relevant codes, we came up with the main themes. We did not compute inter-rater reliability as we conducted thematic analysis on multiple iterations and the themes were developed through refinement of the codes~\cite{mcdonald2019reliability}. 

\subsection{Hypotheses}
Significant research efforts have presented that vocal characteristics of human~\cite{rocklage2018persuasion, wirz2018persuasion} or robot~\cite{kessens2009facial, van2020voice} influence listeners' perceived persuasiveness. Moreover, studies have found perceived persuasiveness to be a significant factor in consumers' purchase decisions in online shopping~\cite{shreffler_persuasiveness_2014}. Vocal characteristics refer to the vocal elements beyond just the words we say, including the tone, personality, gender, and age of the speaker~\cite{poyatos1993paralanguage, marrero2022evaluating}. Therefore, we examined the effect of perceived tone, gender, and age of VA voice on the perceived persuasiveness of a VA and how the persuasiveness ultimately impacts participants' purchase decisions considering the VA's recommendation. We present the following hypotheses based on our research questions.

\textit{\textbf{H1}}: The positive and negative tones of a voice assistant's voice persuade participants more compared to the neutral tone and subsequently, the persuasiveness affects participants' purchase decisions.

\textit{\textbf{H2}}: Varying interactions between the perceived age and gender of a voice assistant's voice persuade participants differently and subsequently, the persuasiveness affects participants' purchase decisions.

\section{Findings}
We used mediation analysis to estimate how perceived tone, age, and gender of the VA voice persuade participants, and subsequently, how the persuasiveness affects participants' purchase decisions. In the following sections, we discuss our findings based on \textit{H1} and \textit{H2}.

\subsection{Effect of Perceived VA tone on VA Voice Persuasiveness and Purchase Decision}

\begin{figure}[!ht]
\centering

  \includegraphics[width=0.95\linewidth]{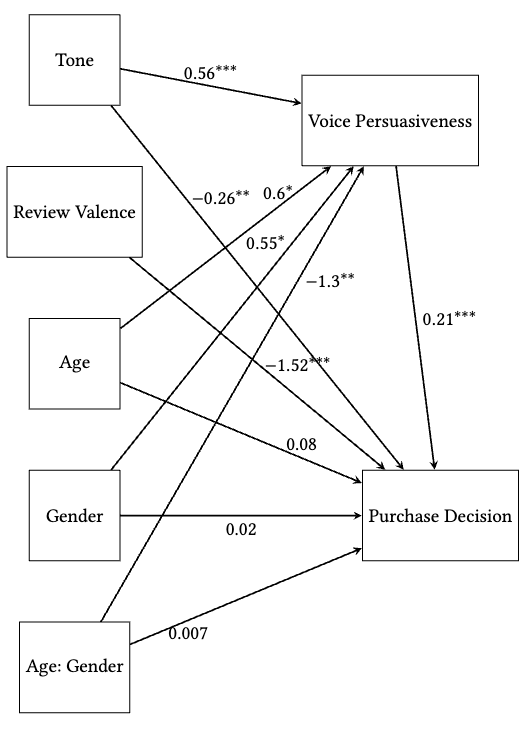}
  \caption{Path model diagram for the mediation of voice persuasiveness in the relationship between purchase decision and the tone, review valence, age, and gender of the VA voice. (* = p < 0.05, ** = p < 0.01, *** = p < 0.001). Baselines for age = younger adult, gender = male}
\label{fig:path-diagram}
\Description{Figure 1: A path diagram shows that the tone (0.56), age (0.6), gender (0.55), and the interaction of age and gender (-1.3) have a significant effect on the VA voice persuasiveness. In addition, tone (-0.26) and review valence (-1.52) have a significant direct effect on the purchase decisions. However, age (0.08), gender (0.02), and the interaction of age and gender (0.007) do not have any significant effect on the purchase decision. Also, VA voice persuasiveness has a significant effect (0.21) on the purchase decisions of the users.}
\end{figure}

\begin{figure}[!ht]
\centering
\begin{subfigure}{0.5\textwidth}
  \centering
  \includegraphics[width=0.95\linewidth]{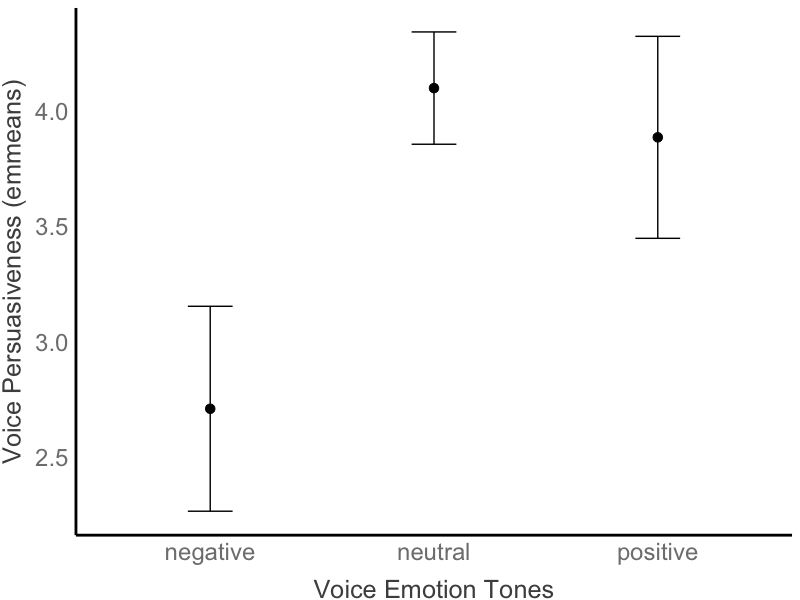}
  \caption{Effects of VA tone}
  \label{fig:voice-persuasive-emotion}
  \Description{Figure 2a: A scatterplot with error bars shows that the participants have a higher perceived persuasiveness of VA voice for neutral and positive emotional valence tones compared to negative-toned voices. The measurements are shown in estimated marginal means values.}
\end{subfigure}%
\\
\begin{subfigure}{0.5\textwidth}
  \centering
  \includegraphics[width=0.95\linewidth]{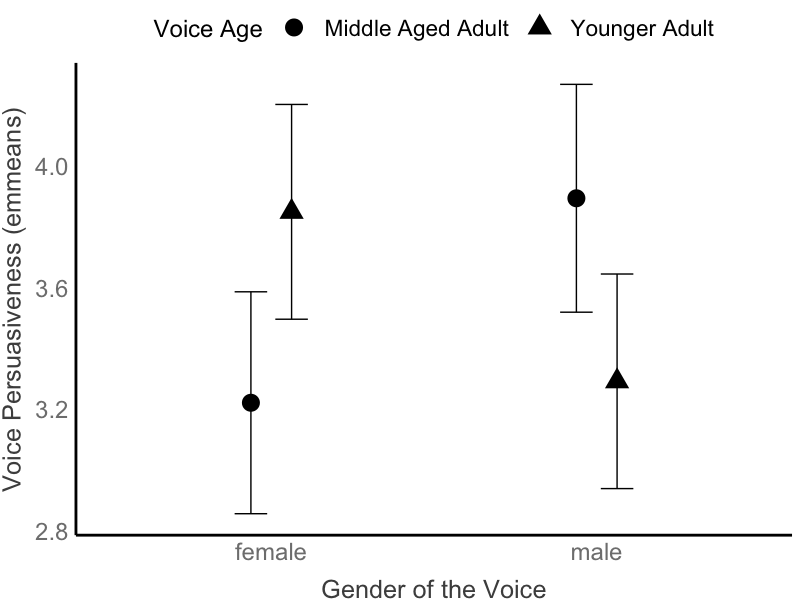}
  \caption{Interaction effects between voice age and gender of the voice }
  \label{fig:voice-persuasive-age-gender}
  \Description{Figure 2b: A scatterplot with error bars shows that the participants have a higher perceived persuasiveness of VA voice for middle-aged male adult voices and younger female adult voices compared to middle-aged female adults and younger male adult voices. The measurements are shown in estimated marginal means values. The values can be found in Table 6a and 6b.}
\end{subfigure}
\caption{Differences in the perceived persuasiveness of voice (estimated marginal means)}
\label{fig:main-effects-persuasion}
\Description{
Figure 2b: A scatterplot with error bars shows that the participants have a higher perceived persuasiveness of VA voice for middle-aged male adult voices and younger female adult voices compared to middle-aged female adults and younger male adult voices. The measurements are shown in estimated marginal means values. The values can be found in Table 3.}
\end{figure}

Table~\ref{tab:effects-mediation} shows that the indirect effect of VA tone on purchase decision was significant, indicating a mediating role of VA voice persuasiveness. This result means that VA tone significantly affected (large effect size) the VA voice persuasiveness, and VA voice persuasiveness in turn significantly affected participants' purchase decisions (Figure~\ref{fig:path-diagram}). A linear regression between VA vocal tone and VA voice persuasiveness revealed that positive (large effect) and neutral tones (medium effect) persuaded participants much more than VA voices with negative tones (Fig. ~\ref{fig:voice-persuasive-emotion}). These results provided partial support for \textit{H1}, where positive and neutral tones had significant effects on persuasiveness and persuasiveness affected users' purchase decisions.

\emph{In summary, our findings suggest that perceived positive and neutral VA tones were significantly more persuasive than negative VA tones, and VA voice persuasiveness significantly affected participants' purchase decisions. In simple terms, neutral VA tones were significantly more persuasive than negative VA tones for negative reviews and ultimately reduced users' purchase likelihood. Concurrently, the positive and neutral VA tones had similar persuasiveness for positive reviews.}

\begin{table}[ht]
  \centering
  \begin{tabular}{ccc}
        \hline
        & Estimate & Std. Error\\
      \hline
       Tone & $0.115^{**}$ & 0.034   \\
       Gender & $0.11^{*}$ & 0.05   \\
       Age & $0.12^{*}$ & 0.05   \\
       Age: Gender & $-0.27^{**}$ & 0.09   \\
     \hline
  \end{tabular}
  \label{tab:anova-summary-persuasiveness-voice}
\newline
\caption{Indirect effect of VA tone and the interaction between VA age and VA gender in the proposed mediation model (* = p < 0.05, ** = p < 0.01, *** = p < 0.001)}
\label{tab:effects-mediation}
\end{table}

\begin{table}[!ht]
    \centering
    \begin{tabular}{cccc}
    \hline
      Voice Age & Voice Gender & Est. & Effect Size  \\
     \hline
     
         Middle Aged & Female < Male & $0.67^*$ & 0.4 (small)    \\
         Middle Aged < Younger & Female & $0.63^*$ & 0.38 (small)    \\
      \hline
    \end{tabular}
    
    \caption{Pairwise comparison of Voice Persuasiveness for the interaction between the gender and age of the voice. Significance after Tukey adjustment is indicated with p-values. Effect size indicates Cohen's \textit{d}. For voice age group, `Middle Aged' is the shorter form for `Middle Aged Adult' and `Younger' is the shorter form of `Younger Adult'.}
  
  \label{tab:pairwise-persuade-gender-age}
\end{table}

\subsection{Effect of Perceived Gender and Age of VA Voice on VA Voice Persuasiveness and Purchase Decision}

We found a significant effect in the interaction of VA voice's perceived age and gender on VA voice persuasiveness (Figure~\ref{fig:path-diagram}). However, we did not find a significant direct effect of the age: gender interaction on participants' purchase decisions. It indicates that there was a full mediation of the voice's persuasiveness between the age: gender interaction and purchase decision of the participants, meaning that age: gender interaction indirectly affected purchase decision through VA voice's persuasiveness. Furthermore, we performed a linear regression between the perceived gender and age of the VA voice and perceived VA voice persuasiveness to determine which combinations of age and gender influenced the VA voice's persuasiveness. We also conducted a post hoc pairwise comparison with Tukey adjustments (Table~\ref{tab:pairwise-persuade-gender-age}). Our findings suggest that participants perceived middle-aged male and younger female voices as more persuasive than younger male voices or middle-aged female voices (small effect size). The effects of the age and gender of the voice are visualized in Figure~\ref{fig:main-effects-persuasion}. These results support \textit{H2} that varying age: gender interactions have various effects on VA persuasiveness, and consequently, the persuasiveness affected participants' purchase decisions.

\emph{Overall analysis suggests that participants found middle-aged male and younger female voices to be more persuasive compared to the other voice types, and the perceived persuasiveness led them to accept the VA recommendation in their purchase decision.}

\subsection{Reasons for Voice Characteristics Preference}

To better understand why our participants preferred certain vocal characteristics, we coded the open-ended responses for additional insights. This helped us comprehend why the participants liked certain voice tones and combinations of ages and genders and were likely to accept suggestions from corresponding VAs. Moreover, this analysis yielded crucial design implications for increased user satisfaction and a user-friendly online shopping interface. 

\subsubsection{\textbf{Positive tone conveyed authenticity}}
A group of participants' reflections about the voice revealed that a positive VA voice created a sense of authenticity.

\begin{quote}
``\textit{The voice felt and sounded authentic and spoke pretty naturally}'' \hfill(P36)
\end{quote}
The genuine feeling came from the comfortable atmosphere caused by the voice assistant's friendly tone. The positive tones of the voices made the listener feel at ease and welcomed.

\begin{quote}
``\textit{The voice assistant had a friendly and inviting quality that made me feel comfortable.}'' \hfill(P157) 
\end{quote}

In a follow-up question, we asked the participants about their personal VA voice preferences. Some participants stated that enthusiastic voices portrayed sincerity in helping out with decision-making. Sometimes participants thought of VAs as sincere salespersons. When the VA sounded keen to help, participants felt that the VA was sincere. Furthermore, multiple participants expressed that flat voices were boring and took away the positive experience from the conversation.
\begin{quote}
``\textit{I want my voice assistant to be high energy and sound like they were happy to help me instead of (displaying) disinterest.}'' \hfill(P290) 

 ``\textit{The (neutral) voice was very flat and monotone. It was boring to listen to at times.}'' \hfill(P89) 
\end{quote}
\subsubsection{\textbf{Neutral tone was reassuring and safe}}
In some cases, participants showed a fondness for neutral voices because of the calmness portrayed by the neutrality of the voice. The flat nature of the tone assured the participants that the VA was not biased in stating the recommendation. Furthermore, a neutral tone created a sense of objectivity which helped participants rely on the recommendation. 

\begin{quote}
\textit{``I like that the voice was neutral sounding; it did not give off any type of extreme emotion, which helped it seem unbiased while it was giving the review of the products.''}  \hfill(P87)

\textit{``I liked that (neutral voice) because I felt like it presented information in an objective manner.'}' \hfill(P148)

\textit{``I liked the (neutral) voice it wasn't too much or too loud, I liked how calm the voice was and it was reassuring.'' }\hfill(P191)
\end{quote}
\subsubsection{\textbf{Positive or negative tone can be overwhelming}}

Some participants expressed their dislike for both positive and negative tones, mentioning that the emotional intensity in these tones can be overwhelming. They preferred calmness because an overenthusiastic voice sounded suspicious (P90). Moreover, some participants suggested that they did not like the negative tone as the tone made it too distracting to focus on the reviews (P146).  

\begin{quote}
``\textit{The (enthusiastic) voice was suspicious that it's selling something}'' \hfill(P90) 

``\textit{The voice with a negative tone sounded uninterested or almost sad. It made it hard to concentrate on the recommendations}.'' \hfill(P146)

``\textit{The voice with negative tone seemed pessimistic towards the majority of things}.'' \hfill(P19)
\end{quote}
Overall, a large number of participants indicated their dislike of positive and negative tones as they wanted to focus more on the content of the reviews and did not want to be tricked. These participants preferred the content and objectivity of the reviews over entertainment.

\subsubsection{\textbf{Middle-aged male voices sounded knowledgeable and experienced}}

A group of participants seemed to trust middle-aged male voices because they felt the voices portrayed confidence, knowledgeability, and reassurance. The confidence portrayed by the low steady pitch of the middle-aged adult voices created a sense of experience as well. As a result, participants perceived the reviews as reliable.
\begin{quote}
``\textit{The voice seemed confident and as if they knew what they were talking about.}'' \hfill(P59)
 
``\textit{It sounded like he had been doing this for a while.}'' \hfill(P111) 
\end{quote}
\subsubsection{\textbf{Younger female voices were soothing}}

A few participants favored younger female voices due to their perceived soothing and appealing quality. The softness in the younger female voice created a soothing ambiance for the participants that triggered a favorable response toward younger female VA voices.

\begin{quote}
``\textit{I liked how the voice was soft and very easy to understand everything that was being said.}'' \hfill(P24)

``\textit{I feel that overall a female voice is more trustworthy and approachable. Her (VA) tone was pleasant and calm}'' \hfill(P278)
\end{quote}
\subsubsection{\textbf{Preference toward familiar or celebrity voice}}

Some participants expressed an interesting intention to use celebrity or familiar voices as VA voices. They thought the familiarity of the voice would create a comfortable and safe ambiance and they would enjoy shopping with a VA having a familiar voice. A few suggestions included `radio DJ from 1989' and `gaming voices'.

\begin{quote}
``\textit{I would create a voice like Ellen DeGeneres, I love everything about her but most of all I love her voice.}'' \hfill(P126)

``\textit{I would use actor voices like Ellen McLain, Morgan Freeman, people that have smooth voices I suppose I like older voices they feel warmer and trustworthy}.'' \hfill(P10)
\end{quote}

\subsubsection{\textbf{Customizable and flexible voice control}}

While articulating their preferences and attitudes toward the voice characteristics of VAs, several participants pointed out that they would have liked different VA voices in varying contexts. Some participants suggested the incorporation of multiple voice characteristics and the flexibility to switch between these voices.

\begin{quote}
``\textit{I would make a voice pack with different emotions and voices that a user could change depending on their mood.}'' \hfill(P20)

``\textit{I think she (the preferred VA) would change her tone depending on the product she was discussing.}'' \hfill(P126)
\end{quote}

\section{Discussion}

We now discuss the implications of our findings and the limitations of the study.

\subsection{Positive and Neutral Tones are More Persuasive and Influence Purchase Decisions}

Our research indicates that both positive and neutral tones in VA voices were perceived as more persuasive than negative tones. Furthermore, this perceived persuasiveness substantially increased the likelihood of participants following the VA recommendation. For instance, participants were persuaded by the VA and were more likely to purchase a product when the VA provided a positive recommendation with positive tones. On the other hand, participants were more persuaded by a neutral tone for negative reviews, and as a result, they were more likely to refrain from purchasing the corresponding product. The positive effect of positive and neutral tone on voice persuasiveness is also evident from the open-ended responses by the participants. Several participants reported that they preferred a positive tone due to the feelings of `comfort' it generated, creating a sense of `authenticity and sincerity.' Likewise, participants found the neutral tone objective and calm. The sense of calmness was reportedly helpful as the participants did not feel forced into making a specific decision, which in turn may have helped the voice appear more persuasive. This outcome can be explained by prior research supporting calm language to be persuasive at a linguistic level~\cite{tan2016winning}. Furthermore, participants indicated an aversion toward the negative tones of VAs, indicating that the negative tones induced negative feelings such as pessimism. This can be explained by prior studies where negative facial expressions and negative activities in a dialogue provoked anxiety in speakers and a neutral or positive attitude from an audience created a sense of comfort~\cite{pertaub2002experiment}.  

The elaboration likelihood model (ELM)~\cite{petty_elaboration_1986} of persuasion states superficial cues such as vocal tone or emotional appeal influence persuasion when people are not motivated to process a message from a communicator deeply. Our model presents evidence that VA tone influenced persuasion, which may indicate participants' lack of motivation or ability to comprehend the content of the reviews. The effect of vocal tone overshadowing the review content might be explained by Delin's argument that the conversational linguistic tone of reviews can make a brand feel `more socially close' to the consumer and ultimately create an attachment with `warm' and `approachable' texts~\cite{delin2005brand}. It is evident that, in the era of VAs' humanlike communication capability, vocal cues may effectively influence users' decision-making.

\subsection{Middle-Aged Male and Younger Female Voices Are More Persuasive and Influence Purchase Decisions}

Our findings indicated that participants attributed higher levels of persuasiveness to the voices of middle-aged males and younger females when compared to other voice types. Because of this increased persuasiveness, they were more likely to follow the VA recommendation for or against product purchase. This result is consistent with participants' open-ended responses on their preference for the VA voice. Participants perceived middle-aged male voices as confident, knowledgeable, and experienced because of their low pitch and calmness. Our findings are consistent with previous research showing that people with lower-pitched voices are more persuasive in various contexts. For example, studies have found that older male voices have a lower pitch than the voices of other age groups or genders. Subsequently, people with lower-pitched voices are more likely to be elected to leadership positions~\cite{klofstad2012sounds}, receive donations~\cite{martin2015effectiveness}, and receive higher ratings of persuasiveness~\cite{mullennix2003social}. 
In contrast, some participants indicated female voices as being more friendly, happier, and soothing. This result is also consistent with the findings of prior studies in which female voices in synthesized form were perceived as more attractive and soothing to participants ~\cite{ernst2020impact, jestin_effects_2022}. 

\subsection{Designing Acceptable and Useful Voice Assistants}

Our findings indicate that participants had varying preferences for vocal characteristics of VAs. This suggests that VA design cannot follow a `one size fits all' approach.  Some participants were persuaded by a more upbeat voice that conveyed a sense of fun and excitement. One group of participants was persuaded by middle-aged male voices which they perceived as experienced and knowledgeable, while another group was persuaded by a younger female voice which they perceived as soothing and youthful. The efficacy of humanlike voice is aligned with the `computers are social actors' (CASA) paradigm~\cite{nass1994computers}. The CASA paradigm states that people often apply social heuristics in their interactions with machines and humanlike activities of a machine may lead to a higher level of engagement. In our study, we also found that some participants find VAs with humanlike voices more engaging and authentic. In contrast, some participants were persuaded by a neutral and flat tone that emphasized factual information over emotional appeals. One underlying reason for this preference might align with the research findings that users sometimes find personalized VA attitudes intrusive and a threat to their privacy~\cite{dev_privacy-preserving_2022}. Furthermore, a few participants flagged varying tones as a `distraction' while listening to the reviews. This preference for a neutral tone can be explained by the central route to persuasion by the elaboration likelihood model (ELM)~\cite{petty_elaboration_1986}. The central route involves a higher level of cognitive processing: keeping the focus on the evidence and logical arguments.

This grouping in preferences might result from participants' inherent motivation. Prior studies have suggested that users motivated by social interaction are likely to perceive a VA as socially attractive and as a friend, whereas users motivated by efficiency are likely to perceive it as an assistant~\cite{choi2021ok}. Differences in user motivation underscore the need to customize VA characteristics. VA designs can align the personality of the VA with that of the user to increase motivation. Several studies in both human-computer interaction (HCI) and psychological fields revealed a significant increase in likability and trustworthiness when a voice matches the user's personality~\cite{braun2019your, snyder2023busting, dahlback2001spoken, dahlback2007similarity}. This aligns with the \textit{similarity-attraction} theory: people are more likely to be attracted to those who are like them~\cite{byrne1967attraction}. Another avenue for VA personalization lies in enhancing accessibility for marginalized and vulnerable populations. For example, older adults may favor the voices of middle-aged adults because they are generally easier to understand and they speak at a slower rate than younger adults~\cite{winkler2007influences}.

In addition to personalized voices, many participants favored VAs with the capability to change vocal expression based on a situation or a function. Desai and Twidale demonstrated that users might perceive virtual assistants (VAs) as different personas depending on the context, using various metaphors to interact with VAs accordingly~\cite{10.1145/3609326}. Adapting VA tones to the situation can enhance the user's impression. The preference for situation-aware voices is aligned with previous findings that users prefer personalized conversation by chatbots based on varying functions~\cite{volkel_personalised_2020}. Such adaptability could be beneficial in specific scenarios. For example, the use of a warm vocal tone can be an effective way to mitigate user disappointment when a VA fails to communicate with the user as expected~\cite{huang2023should}. Furthermore, the ability of a VA to reflect multiple personalities can evoke the social presence heuristic~\cite{lee2004multiple} which suggests that users are more likely to believe and trust a VA~\cite{kim2016interacting}. In addition, the ability to easily change VA tone based on the mood might be a delightful experience for the users. 

The current versions of voice assistants such as Alexa, Google Nest, or Siri provide limited variation in the voices, encompassing only gender or accent-based variations~\cite{alexawhisper, googlecolor}. Our study indicates that introducing variations in tone or age for VA voices could yield a more personalized and engaging user experience in VA usage, particularly enhancing trust and comfort during complex tasks such as online shopping. Furthermore, as technology advances, there is a potential for VAs to create completely realistic and credible speech, incorporating natural intonations and emotions according to appropriate context~\cite{noufi_role_2023, pamisetty_prosody-tts_2023}. These enhancements of VA voices can facilitate more VA engagement, ultimately increasing VA adoption and providing users with a more seamless and intuitive experience, particularly beneficial for disabled and older adult individuals seeking hands-free alternatives. However, alongside the advancement of VA voices, it is imperative to consider the ethical ramifications of creating AI voices indistinguishable from human voices to prevent their potential misuse for deception or manipulation of users. Therefore, future studies should not only focus on advancing the technology to make VA voices humanlike and context-aware but also ethically evaluate the consequences and potential misuse of such advancements and formulate intelligent policies to mitigate the risk of manipulative VA voices.

\subsection{Influence of Cultural and Linguistic Backgrounds}
Currently, voice assistants speak multiple languages for users from different linguistic backgrounds~\cite{noauthor_how_2020}. Studies have shown the effect of culturally specific principles in voice emotion recognition---affecting the perception of the valence of vocal tones in different languages~\cite{paulmann2014cross, yang_take_2021}.  Our study evaluated American participants' perceived tones of the VAs. Hence, the findings of this study should be interpreted carefully considering the cultural background of the participants, as the perception of tone or age in a VA voice may vary in different cultural and linguistic contexts. For example, research has shown that Japanese people pay more attention to the emotional tones in voices compared to Dutch people~\cite{tanaka_i_2010}. Also, there are differences in how native Mandarin and native American speakers understand emotions in voice tones~\cite{chen_cultural_2023}. Furthermore, users from Japanese, Mandarin, and Brazilian Portuguese backgrounds were found to have different perceptions of the positivity and negativity of the vocal tones in the same linguistic contents~\cite{erickson_cross_2020}. Therefore, future research should consider the cultural backgrounds of VA users and compare VAs with different languages to understand how people from different backgrounds comprehend the tone or age of the voice.

\subsection{Ethical Consideration and Healthy Persuasion}

Although our findings suggest the effectiveness of VA vocal characteristics to influence users' purchase decisions, safeguarding user interest is an important aspect of VA designs. Ethical considerations, including the permissible degree of persuasive tactics and mechanisms to ensure informed consumer decisions, are critical. Moreover, it is crucial to distinguish between helping users make good decisions (healthy persuasion) and deceiving them (manipulation)~\cite{harre1985persuasion, noggle_manipulation_2018}. This is especially true for special interest groups susceptible to manipulative practices, including older adults or individuals with visual impairments~\cite{kemp_consumer_2023}. Tailoring personalized VA voices fosters inclusivity and empowers people with visual impairment or older adults to utilize hands-free technologies with trust and confidence. However, these vulnerable demographics have limited resources to verify the source of new information, often relying on other people for assistance~\cite{akter_i_2020}. In addition, potential cognitive or physical limitations may limit their capability to discern manipulative tactics, making them a potential target for exploitation through manipulation~\cite{shang_psychology_2022}. One strategy to mitigate the risk of manipulation is to provide comprehensible disclaimers and offer adequate training. Moreover, utilizing automated mechanisms to identify voice-based manipulation could be valuable in preventing deceptive practices via VAs. Furthermore, research found that a subset of older adults prefers robots to exhibit robotic characteristics rather than human-like qualities~\cite{berridge_companion_2023}, aligning with our study's findings that some participants favor robotic voices in VAs as a reminder of the machine nature of the technology. Therefore, VA designs should prioritize user preferences, particularly those of vulnerable populations, before implementing alterations to VA voices~\cite{oumard_pardon_2022}.

We also need to be careful how we interpret the findings related to preference for a particular gendered voice. Previous research has observed a tendency toward gender stereotypes in preferences for synthesized voices~\cite{dogruel_gender_2023}. However, the result of this study may not be applicable in different contexts or with different voice stimuli. Therefore, it is crucial to ensure that voice assistant design does not inadvertently perpetuate gender stereotypes. Rather, the VA design should focus on enhancing user comfort and provide gendered or gender-ambiguous voice options~\cite{tolmeijer_female_2021} as a personalization alternative. Moreover, letting users easily change the VA voice can protect their interests. This way, users can choose voices that make them feel comfortable and enjoy shopping  more. By integrating brief disclaimers and easy customization, VA designs respect users' choices and make sure they have a good experience without feeling pressured.

\subsection{Limitation and Future Work}

There are a few limitations in this study. First, the voice dataset we used consisted of only male and female voices, as there are limited resources on generating gender-ambiguous~\cite{sutton_gender_2020} voices with the desired tones and age groups. Significant efforts have been made toward making VAs more inclusive, with gender-ambiguous text-to-speech (TTS) systems that can produce non-binary voices~\cite{danielescu2023creating}. However, a considerable proportion of users are yet to adopt the concept of a gender-ambiguous voice~\cite{mooshammer_gender_2022}. Future work should incorporate variations in the vocal tone and age of gender-ambiguous voices to yield more precise results free of gender stereotypical perceptions.

Furthermore, we conducted the study in an online platform with synthesized voices to minimize variability in the simulated scenarios. However, this study design did not allow researchers to control environmental factors such as the volume of the voices or the participants' surroundings. Moreover, this approach did not provide participants with an authentic VA interaction experience involving a physical VA and the inherent motivation of users to engage in product shopping. Although we curated the most popular gender-neutral products suitable for online shopping with voice-assisted reviews, ensuring minimal visual inspection, the limited selection of predefined product categories may not fully capture responses across diverse product types. Therefore, there is a possibility that participants' responses may differ in a real-world VA interaction setting. Future work should evaluate the feasibility of conducting VA experiments in a controlled laboratory setting with a physical VA and a broader array of product categories.

Moreover, we did not explore the distinction between persuasion and manipulation, despite their correlation and overlap in scope. Similar research on persuasiveness should also consider the effects of manipulation.

\section{Conclusion}

As voice assistants (VAs) become more capable of assisting people in complex tasks, it is essential to identify the factors that enhance their persuasiveness and the strategies through which VAs can engender user trust, thereby enhancing reliance on VA-provided recommendations. Our findings suggest that voice characteristics play an important role in shaping user decisions in online shopping with VAs. Participants found neutral and positive-toned simulated voices more persuasive than negative-toned voices and tended to follow positive and neutral-toned recommendations more often. Additionally, middle-aged male and younger female voices were perceived as more persuasive, prompting users to more frequently follow recommendations delivered in these vocal tones.  Our findings suggest that varying voice characteristics can create a sense of comfort and authenticity among users, improving user adoption of VAs in tasks with financial consequences such as online shopping. However, ethical considerations, such as implementing disclaimers regarding the VA's vocal tone or measures to prevent the misuse of vocal characteristics for deceptive purposes, need to be reinforced in VA design to differentiate legitimate persuasion from manipulation. Moreover, it is imperative to exercise caution before generalizing findings related to gender stereotypical perceptions to avoid oversimplification and further amplification of gender biases.

\begin{acks}
   This material is based upon work supported by the National Science Foundation under grant CNS-2207019 and the Social Science Research Funding Program (SSRFP) at Indiana University. We would also like to thank Hannah Bolte and Elizabeth Ray at the Indiana Statistical Council Center (ISCC) for their guidance on statistics analysis. 
\end{acks}

\bibliographystyle{ACM-Reference-Format}
\bibliography{ref-auth1, references}

\clearpage
\appendix

\begin{table}[ht!]
  \begin{tabular}{lp{4.5cm}}
        \hline
      Factor & Item \\
     \hline
       VA Usage & How often do you interact with a Voice Assistant?  \\
       \hline
      \multirow{4}{*}{Disposition of Trust~\cite{gefen2000commerce}} & I generally trust other people    \\
       & I tend to count upon other people  \\
       & I generally have faith in humanity   \\
       & I generally trust other people unless they give me a reason not to    \\
       \hline 
  \end{tabular}
  \caption{Question Items about usage of voice assistant and disposition of trust}
  \label{tab:ques-sva-usage-trust}
\end{table}

\begin{table}[ht]
  \begin{tabular}{llll}
        \hline
       & Positive & Negative & Neutral \\
     \hline
       \multirow{2}{*}{Younger Male} & 204.7 & 120.7 & 121.5  \\
       & (6.7) & (5.8) & (6.5) \\
       \hline
      \multirow{2}{*}{Younger Female} & 258.2 & 203.4 & 209.7 \\
      & (6.5) & (5.3) & (6.4)  \\
       \hline
       \multirow{2}{*}{Middle-Aged Male} & 155 & 125.9 & 129 \\
       & (6) & (5.4) & (5.4) \\
       \hline
       \multirow{2}{*}{Middle-Aged Female}  & 247.8 & 164.8 & 182.7 \\
       & (5.9) & (5.1) & (5.6) \\
       \hline
  \end{tabular}
  \caption{Mean pitch (Hz) and speech rate (in parenthesis) of tone, gender, and age combinations for the synthesized speech in the study}
  \label{tab:pitch-sr}
\end{table}

\begin{table}[ht!]
\begin{tabular}{lll}
\hline
  Categories & \multicolumn{2}{c}{Frequency}\\
 \hline
 
  & Stimulus Validation & Main Study\\
 \hline
 \multicolumn{3}{c}{\textbf{Gender}}\\
 \hline
   Female & 38 (48.7\%) & 161 (48.1\%) \\
   Male & 38 (48.7\%) & 168 (50.1\%) \\
   Non-binary & 2 (2.5\%) & 5 (1.5\%) \\
   Others & 0 (0.0\%) & 1 (0.3\%) \\
 \hline
 \multicolumn{3}{c}{\textbf{Age Group}}\\
 \hline
   18-29 & 36 (46.1\%) & 135 (40.3\%) \\
   30-49 & 31 (39.7\%) &  144 (43.0\%)  \\
   50-64 & 10 (12.8\%) &  46 (13.7\%) \\
   65+ & 1 (1.3\%) &  10 (3.0\%) \\
   \hline
 \multicolumn{3}{c}{\textbf{Highest Education Level}}\\
  \hline
   No high school & 3 (3.8\%) & 2 (0.6\%) \\
   High school or equivalent & 34 (43.6\%) & 63 (18.8\%) \\
   Undergrad.\slash 2 year degree & 35 (44.8\%) & 227 (67.75\%) \\
   Graduate degree & 6 (7.7\%) & 43 (12.8\%)  \\
   \hline
 \multicolumn{3}{c}{\textbf{Racial or Ethnic Background}}\\
 \hline
   Hispanic or Latino & 5 (6.4\%) &  25 (7.5\%)\\
   Black or African American & 7 (9.0\%) &  31 (9.25\%)\\
   American Indian & 0 (0\%) & 4 (1.2\%)\\
   Asian & 7 (9.0\%) &  26 (7.8\%)\\
   White & 54 (69.2\%) & 222 (66.3\%)\\
   Mixed & 4 (5.1\%) & 25 (7.5\%)\\
   Other & 1 (1.3\%) & 2 (6.0\%)\\
   \hline
 \multicolumn{3}{c}{\textbf{Using Voice Assistant}}\\
  \hline
   Never & 8 (10.4\%) &  19 (5.7\%)\\
   Less than once a month & 10 (13.0\%) &  39 (11.6\%)\\
   Once a month & 5 (6.5\%) & 24 (7.2\%)\\
   Multiple times a month & 13 (16.9\%) &  70 (20.9\%)\\
   Once a week & 2 (2.6\%) & 29 (8.7\%)\\
   Multiple times a week & 14 (18.2\%) & 75 (22.4\%)\\
   Once a day & 5 (6.5\%) & 14 (4.2\%)\\
   Multiple times a day & 20 (26.0\%) & 65 (19.4\%)\\
  \hline
    & N= 78 & N= 335\\
  \hline
\end{tabular}
\caption{Demographic and voice assistant usage information of final participants}
\label{tab:demography}
\end{table}

\end{document}